\documentclass[twocolumn,showpacs,preprintnumbers,amsmath,amssymb]{revtex4}

\usepackage{graphicx}
\usepackage{dcolumn}
\usepackage{bm}

\begin{document}

\title{Evidence for the role of normal-state electrons in nanoelectromechanical damping mechanisms at very low temperatures}
\author{K.J. Lulla, M. Defoort, C. Blanc, O. Bourgeois, and E. Collin}

\address{
Institut N\'eel
\\
CNRS et Universit\'e Joseph Fourier, \\
BP 166, 38042 Grenoble Cedex 9, France \\
}

\date{\today}

\begin{abstract}
We report on experiments performed at low temperatures on aluminum covered silicon nanoelectromechanical resonators.
The substantial difference observed between the mechanical dissipation in the normal and superconducting states measured within the same device unambiguously demonstrates the importance of normal-state electrons in the damping mechanism. The dissipative component becomes vanishingly small at very low temperatures in the superconducting state, leading to exceptional values for the quality factor of such small silicon structures. A critical discussion is given within the framework of the standard tunneling model.  
\end{abstract}


\maketitle
%
%

Micro and nanomechanical devices are under intense investigation for both their promising instrumental applications and their implication in fundamental issues of physics. 
These devices are ultra-sensitive mass \cite{mass} and force detectors \cite{force}, they can be used in their linear \cite{lin} or nonlinear regimes \cite{westra} to implement various signal processing schemes \cite{bifurc,mohantyNature}.
In a more fundamental realm, they can be thought of as probes for non-newtonian deviations to gravity at small scales \cite{gravit}, for refined studies of the Casimir force \cite{casimir}, and for the study of quantum fluids \cite{physicaBus}.
Moreover, nanoresonators themselves cooled to their quantum ground state tackle problems that have been around quantum mechanics since the early beginning, with the possibility of controlling a mechanical collective macroscopic degree of freedom at the quantum level \cite{cleland,simmonds,lehnert,schwabNanoQal}.

Having high quality devices is desirable in many of these fields. 
However, it is well known that the quality factor $Q$ of mechanical structures becomes worse as their size is reduced \cite{roukesdamping2002},
while internal stresses have been found to drastically increase the $Q$ in silicon-nitride nanobeams \cite{jeevakSiN}.
Although it is clear that the surface-to-volume ratio is a key ingredient for the understanding of mechanical dissipation, a proper theoretical explanation covering all experiments remains elusive \cite{siliconbeam,kunalthesis,kunalPRB,pashkin}.
Nanomechanical friction mechanisms thus deserve to be understood from both an engineering and a fundamental condensed matter physics point of view.

Almost all nanoresonators used in dissipation experiments possess a metallic coating used to actuate and detect the motion. This layer has an essential impact on the mechanical properties, since it adds mass and surface stresses which significantly modify the dissipation characteristics \cite{kunalthesis,Parpiaoscill}. Most experiments are performed with normal conducting metals; only little is known about superconductor-covered nanodevices \cite{schwabNanoQal,JaredPhd,mikka}. 

Addressing dissipation mechanisms requires a broad temperature range to be explored, within the Kelvin and sub-Kelvin range.
Common features are observed: the dissipation follows a power law $T^n$ below a certain temperature $T^*$, with a crossover to a rather flat high temperature region that depends on the nature and size of the object. The resonance frequency shifts logarithmically at the lowest temperatures, and reveals a maximum around the same crossover temperature $T^*$. These features are commonly attributed to tunneling Two-Level Systems (TLS) present in the devices, mimicking the results obtained on bulk (amorphous) dielectric and metallic materials \cite{esquinaziTLS,parpiaSiO,enss}.
However, no experimental consensus exists with respect to the exponent $n$, and fits to the Standard Tunneling Model (STM) usually fail to be quantitatively consistent \cite{siliconbeam,kunalthesis,kunalPRB,pashkin}.

In the present paper we report on experiments performed on aluminum-covered goalpost silicon nanodevices resonating around 7$~$MHz \cite{usJLTPQFS}. 
The goalpost structure consists of two feet 3$~\mu$m long linked by a 7$~\mu$m paddle, all about 250$\times$150$~$nm wide and thick. The aluminum layer is 30$~$nm thick, with a superconducting $T_c$ of 1.55$~$K \cite{TcAl}.
The upper critical field of our nanowires is around 0.5$~$T \cite{SupraFilm,AlHc}, and the corresponding critical current is 40$~\mu$A.
The sample is placed at low temperatures in cryogenic vacuum (P $ < 10^{-6}~$mbar). The motion $x(t)$ is actuated and detected using the magnetomotive scheme \cite{clelandMagneto}. The resonator has been fully characterized and calibrated in its cryogenic environment \cite{calibpaper}; care has been taken to minimize impedance loading losses due to circuitry. Joule heating in these 100$~\Omega$ devices has been carefully characterized. 


We have measured the frequency and resonance linewidth of the first out-of-plane flexural mode of the structure from 30$~$K down to about 35$~$mK.
The 4.2$~$K quality factor $Q$ is about $5~000$, consistent with the literature \cite{roukesdamping2002}.
Measurements have been performed at low in-plane static fields $B$ and small sinusoidal currents $I_0$, 
enabling to probe the normal ({\bf N}) and superconducting ({\bf S}) states of the aluminum coating.
Details on the Methods can be found in the Supplemental Material \cite{supplemental}.
\newpage

\begin{figure}
\includegraphics[height=6.5 cm]{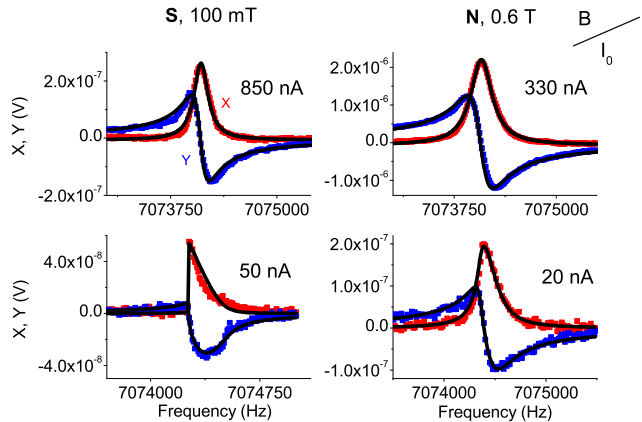}
\caption{\label{figure1} (Color online) Mechanical resonance lines measured in vacuum at 100$~$mK (raw data from lock-in detector, in-phase X and out-of-phase Y components). Magnetomotive settings $B,I_0$ given within the graph (current in {\it rms} units). {\bf S} stands for superconducting state, and { \bf N} for normal. Full lines are fits; note the nonlinear shape of the low field and low current resonance in {\bf S} state (see text).}
\end{figure}

Typical resonance lines obtained at 100$~$mK are shown in Fig. \ref{figure1}. The in-phase (X) and out-of-phase (Y) components obtained from the lock-in measurement are displayed. 
A key feature of the data is that in the superconducting state, the mechanical resonance becomes nonlinear (bottom left graph in Fig. \ref{figure1}). 
Extracting the intrinsic damping properties of the resonator requires thus to carefully describe these nonlinear phenomena.

The overall lineshape can be captured within a single Duffing frequency-pulling term $\beta(B,I_0)$ that depends on both the magnetic field and drive current amplitude. 
For a high-$Q$ oscillator, this term is purely dispersive and changes the resonance frequency $f$ of the device through $f=f_0+ \beta\, x^2$, with no impact on the friction.
This term happens to be also temperature-dependent, and greater at lower field/current excitations (see Supplemental Material \cite{supplemental}). 
Duffing-like nonlinearities recall experiments performed on larger superconducting structures (vibrating wires) \cite{KonigJLTP}.
We suggest that their origin lies within the dynamics of the superconducting vortex state.


The damping is described by a friction force $F_{damp.}=-2 \Lambda \, \dot{x}$ with $\dot{x}$ the speed of the moving structure.
The damping coefficient $\Lambda$ can be converted in units of Hz through $\Delta f = \frac{1}{2 \pi} 2 \Lambda/m$ with $m$ the mass of the oscillator. In the linear regime $\Delta f$ is the full width at half height of the resonance (FWHH).
The measured damping happens to be also nonlinear. We write $\Lambda(x) = \Lambda_0 + \Lambda' \left| x \right|$ with $\Lambda'(B,I_0)$ a temperature-dependent friction nonlinear coefficient that captures the observed behaviors, and $\left| x \right|$ the amplitude of the motion. The quoted dampings correspond to $\Delta f (x_{max})$, with $x_{max}$ the motion amplitude at the peak of the resonance.

\vspace{0.2cm}

\vspace{1mm}
\baselineskip 12pt

\begin{figure}
\includegraphics[height=5.75 cm]{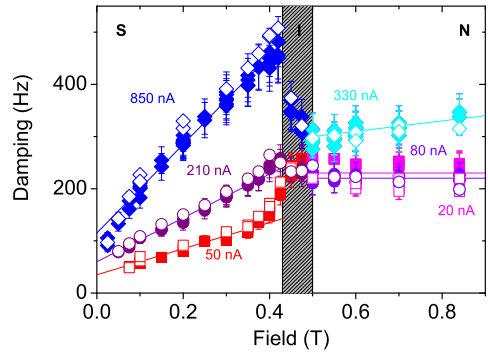}
\caption{\label{figure2} (Color online) Nonlinear damping as a function of magnetic field at 100$~$mK. 
Three driving current settings are displayed, for both {\bf S} and {\bf N} states. The shaded area corresponds to an intermediate range where both states coexist ({\bf I}). Empty symbols are obtained from complete nonlinear fits of the resonance lineshapes; full symbols are from the linear range or recalculated heights (see text). 
Lines are linear guides through the data.}
\end{figure}

\begin{figure}
\includegraphics[height=6.5 cm]{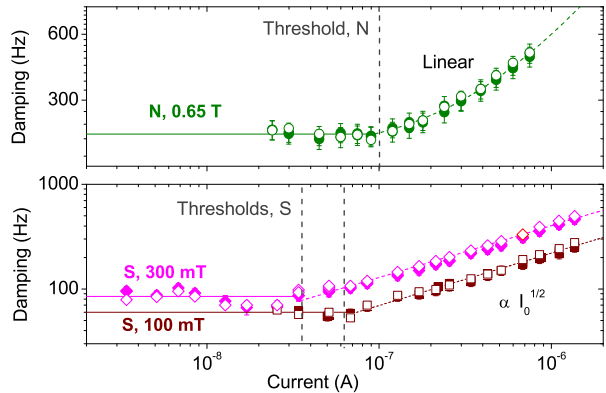}
\caption{\label{figure3}
(Color online) Nonlinear damping as a function of drive current at 100$~$mK. Three fields are displayed (with two leading to the {\bf S} state, below, and the other one to the {\bf N} state, above). Empty and full symbols follow the same nomenclature as Fig. \ref{figure2}. In each state, a threshold effect is seen (dashed verticals), followed by different growths (linear or $I_0^{1/2}$ guide to the eye). Note the log.-log. scale. 
}
\end{figure}

In Fig. \ref{figure2} we show the dissipation coefficient obtained as a function of the magnetic field for three current settings (data taken at 100$~$mK). Normal and superconducting regions are clearly identified, with a coexistence zone in between ({\bf I}).
In Fig. \ref{figure3} the data are presented as a function of the drive current, for three magnetic fields (two in the {\bf S} and the other one in the {\bf N} states). 
When the lineshape becomes too nonlinear, we can still recompute the nonlinear FWHH from the height of the resonance peak, making sure the frequency sweep has been performed in the proper upwards/downwards direction with respect to $\beta$ \cite{PRBDuffing}. FWHH obtained from "brute force" fits of the nonlinear lines are also displayed (open symbols; see Supplemental Material \cite{supplemental}).
Note that the measurements presented are free of impedance loading losses $\propto B^2$ \cite{clelandMagneto}; only intrinsic dependencies are displayed.  
The behavior in the {\bf N} and {\bf S} states is drastically different.

\vspace{0.2cm}

\vspace{1mm}
\baselineskip 12pt

In the normal state, acoustic damping properties are known to be strain-dependent \cite{ramos2000}.
For goalpost mechanical devices, the linewidth is found to be constant until a threshold and then grows linearly with respect to the displacement $\left|x\right|$, a feature that is characteristic of the inelasticity of the metallic layer \cite{metalcoatings,parametricPRB}. 

On the other hand, below 0.45$~$T in the superconducting state a drastically different nonlinear damping mechanism is visible. Above a certain  displacement threshold, the damping seems to follow $I_0^{1/2}$ (Fig. \ref{figure3}). At the same time, both the plateau value before threshold and the prefactor of the $I_0^{1/2}$ law after threshold seem to depend linearly on the magnetic field, as can be seen with the straight lines in Fig. \ref{figure2}. Only close to the critical field at low currents, or close to zero field at large currents, departures from this simple linear-in-$B$ description can be noticed (see Supplemental Material \cite{supplemental}). 
Fig. \ref{figure2} recalls superconducting nonlinear damping results understood in terms of vortex dynamics obtained on much larger reeds \cite{brandt}. As for $\beta$, the vortex state in the superconducting layer should be the source of this behavior through $\Lambda'$. 
These nonlinear results will be published elsewere.

\begin{figure}
\includegraphics[height=6.3 cm]{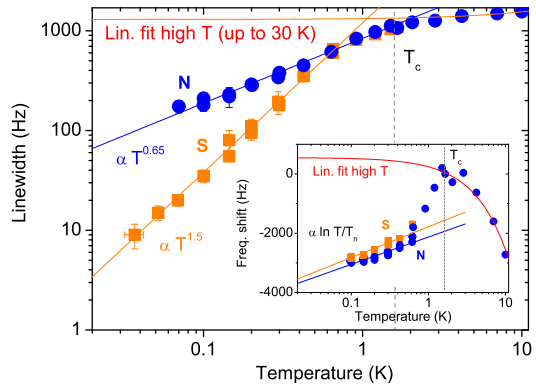}
\caption{\label{figure4} (Color online) Resonance linewidth $\frac{1}{2 \pi} 2 \Lambda_0/m$ (main) and frequency shift (inset) in the zero-drive limit ($B \rightarrow 0$ and $I_0 \rightarrow 0$) as a function of temperature. The zero of frequency shifts has been arbitrarily chosen at 2$~$K (maximum). The high temperature empirical linear fits \cite{calibpaper} (valid up to 30$~$K) are shown, with the dashed vertical corresponding to the metallic layer $T_c$. Power law fits and logarithmic functions are displayed (main graph and inset respectively).}
\end{figure}

All the features we observe are robust and seen at all temperatures below typically $T_c/2$, with stronger expression as $T$ is reduced. 
Knowing the nonlinear behaviors of Fig. \ref{figure2} and Fig. \ref{figure3} we extract the intrinsic, linear damping properties.
In Fig. \ref{figure4} we present data obtained after carefully extrapolating the measurements at low fields and low currents, for each measured temperature, ramping up from 35$~$mK. Radio-frequency over-heating was present due to imperfect filtering, which caused the normal-state measurements to saturate below about 80$~$mK (not shown).
On the other hand, the superconducting state linewidth continues to fall down as $T$ is decreased.
The reproducibility of the results has been checked on several cool-downs from 300$~$K, for two samples of the same chip. 

\vspace{0.2cm}

\vspace{1mm}
\baselineskip 12pt


For our device, the cross-over $T^*$ between low and high temperature behavior lies around 2$~$K \cite{siliconbeam,kunalPRB,pashkin}. 
Above $T^*$, both linewidth and frequency shift display linear temperature dependencies, and   
we reproduce in Fig. \ref{figure4} the empirical fits of Ref. \cite{calibpaper}.
Normal and superconducting curves split around 700$~$mK, which is about the temperature where the fraction of normal state electrons $\rho_n$ in the superconductor starts to substantially decrease, around $T_c/2$ \cite{SupraBook}.
However, the dissipation (FWHH) in the {\bf S} state does not show exponential behavior: for both {\bf S} and {\bf N} it follows different power laws $T^n$ while the frequency shifts display the accepted logarithmic tendency (see inset, Fig. \ref{figure4}). 
This is in drastic contrast with measurements performed on micromechanical devices having thicker metallic coatings \cite{jeevakAlandAg}, and micron-sized vibrating wires \cite{esquinazi} which do not show any difference between normal and superconducting states.

\vspace{0.2cm}

\vspace{1mm}
\baselineskip 12pt

Fig. \ref{figure4} tells us that the electrons play a fundamental role in the mechanical dissipation mechanism. 
The standard discussion invokes the Standard Tunneling Model (STM) \cite{esquinaziTLS}. In this model, TLS present in the structure (which exact nature is not known) are coupled to the strain field induced by the motion of the mechanical mode \cite{miles}. They can absorb energy from the mode, which they release to the outside world through their coupling to phonons and/or electrons. 
For the materials in use here (Al and Si), the dominant phonon wavelength $\lambda_{dom}= h v_{ph}/(2.82 k_B T)$ is of the order of the transverse dimensions of the feet of the structures around $1~$K (with $v_{ph}$ the speed of sound). This means that phonons will gradually cross-over from a 3D to a 1D regime as the temperature is reduced. On the other hand, electrons remain 3D since the Fermi wavelength $\lambda_F$ in the metal is smaller than a nanometer.
Depending on the dispersion relation of the mechanical mode ($\omega \propto k$ or $\omega \propto k^2$), electron and phonon mechanisms lead to different low temperature dissipative power law behaviors $T^n$ (accompanied by logarithmic frequency shifts). 
For a string vibration with $\omega \propto k$, Ref. \cite{pashkin} finds a $T^1$ law for normal-state aluminum nanowires. Their conclusions lead to a phononic mechanism, consistent with experiments performed on superconducting aluminum nanowires which display also a linear temperature dependence \cite{mikka}. 
Our finding on aluminum-covered silicon cantilevers is drastically different, and closer (in the normal state) to the gold nanowire result $T^{0.5\pm0.05}$ of Ref. \cite{kunalPRB}.
For pure flexure $\omega \propto k^2$, a modified version of the STM for surface TLS coupled to phonons predicts $T^{1/2}$ \cite{castro}. 
This is rather close to our finding $T^{0.65\pm0.1}$, but as we already said the proper mechanism applying to our experiments has to involve electrons. We believe that an extension of the TLS-electron model in the presence of a $\omega \propto k^2$ flexural mode should also lead to a $T^{1/2}$ dependence. 
In the TLS-electron picture, when superconductivity occurs the damping falls down rather rapidly with temperature \cite{haust,WeissPhysicaB}. This is qualitatively consistent with our data, but does not reproduce the $T^{1.5\pm0.1}$ dependence of Fig. \ref{figure4}. One could conjecture that phonon-mediated dissipation is much smaller than a superconductivity-induced damping that dominates and decreases below 700$~$mK. 
As a conclusion, to our knowledge no published extension of the TLS model is able to reproduce our results.
Note that experimentally, the main difference between the aluminum nanowires and our experiments is the presence/absence of stress which changes the dispersion law of the resonance mode, and potentially the strength of the different dissipation mechanisms.

\vspace{1mm}
\baselineskip 12pt

In summary, we measured experimentally at millikelvin temperatures the mechanical damping of aluminum-plated silicon nanoelectromechanical devices resonating around 7$~$MHz.
The experiment has been performed in both the normal and the superconducting states of the metallic coating. We found a striking difference between the two states, proving that normal-state electrons play a key role in the nanomechanical dissipation mechanism. 
This has never been reported before.
Power law dependencies with respect to temperature have been found for the dissipation, with different exponents for {\bf S} and {\bf N} states.
A complex nonlinear behavior in the superconducting state has been identified, which we believe is due to the dynamics of the vortex state. 
In the superconducting state, the quality factor $Q$ reaches about a million below 40$~$mK. 
Our results shine a new light on other experimental data available that still cannot be explained in a global and consistent way by a single model. Low-temperature nanomechanical damping remains a challenge for physicists, and the present work is clearly calling for new theoretical input.  

The authors want to greatfully thank H. Godfrin and A. Armour for extremely valuable discussions, and J. Minet and C. Guttin for help in setting up the experiment. 
We wish to thank T. Fournier and T. Crozes for help in the microfabrication process. We acknowledge the support from MICROKELVIN, the EU FRP7 low temperature infrastructure grant 228464 and of the 2010 ANR French grant QNM n$^\circ$ 0404 01.

\end{document}